\documentclass[epsfile,prc,aps,showpacs]{revtex4}
\usepackage{graphicx}
\usepackage{epsfig}

\begin{document}

\title{Nuclear surface studies with  antiprotonic atom X-rays }
\author{S. Wycech$^1$,
F.J. Hartmann$^2$, J. Jastrz\c{e}bski$^3$,  B. K\l os$^4$, A.
Trzci{\'n}ska$^3$, T. von Egidy$^2$ }

\affiliation{$^1$ So{\l}tan Institute for Nuclear Studies, Warsaw,
Poland}

\affiliation{$^2$ Physics-Department, Technical University of
Munich, Garching, Germany }

\affiliation{$^3$ University of Warsaw, Warsaw, Poland}
\affiliation{$^4$University of Silesia, Katowice, Poland}

\date{\today}

\begin{abstract}
  The recent and older level shifts and widths in $\overline{p}$ atoms
  are analyzed. The results are fitted by an antiproton-nucleus
  optical potential with two basic complex strength parameters. These
  parameters are related to average $S$ and $P$ wave scattering
  parameters in the subthreshold energy region. A fair consistency of
  the X-ray data for all $Z$ values, stopping $\overline{p}$ data and
  the $\overline{N}N$ scattering data has been achieved. The
  determination of neutron density profiles at the nuclear surface is
  undertaken, and the determination of the neutron $ R_{rms}$ radii is
  attempted. Uncertainties due to the input data and the procedure are
  discussed.
\end{abstract}


\maketitle

\section {Introduction}

Recent CERN/LEAR measurements of X-ray transitions in antiprotonic
atoms provide new data on the level shifts and widths in heavy
systems~\cite{TRZ01}. These results complement older CERN, BNL and
other~\cite{ROB77,KOH86,ROH86} studies in medium Z atoms and fairly
recent experiments with hydrogen, deuterium and
helium~\cite{AUG99,AUG99a,GOT99,SCH91}. The main distinction of the
new results from those obtained earlier in light nuclei is a much
broader data basis, extended to the region of neutron excess and
deformed nuclei. With the enlarged data one can attempt an extraction
of a phenomenological $\bar{p}$ optical potential that contains
separate strengths of $p\overline{p}$ and $n\overline{p}$
interactions~\cite{BAT97,FRI05}. It allows to study properties of the
neutron density distributions at the nuclear surface. As stressed on
many occasions the $\overline{p}$ atoms are well suited for such
studies. A problem is met on this way as the best description of the
atomic data indicate the $p\overline{p}$ and $n\overline{p}$
absorption rates to be roughly the same. This result is in conflict
with the chamber studies of low energy $\overline{p}$
annihilation~\cite{BIZ74,BAL89}.  The latter experiments indicate the
$p\overline{p}$ absorption rate to be twice as fast as the
$n\overline{p}$ one. This discrepancy has to be understood and we
indicate the solution.

The content of this paper is as follows. In section II, the optical
potential is constructed for $\overline{p}$ bound into atomic
states. It consists of two terms, the central and the gradient one.
The strengths of these terms are phenomenological parameters which
describe the $S$ and $P$ wave $\overline{N}N$ scattering amplitudes
extrapolated below the $\overline{N}N $ threshold. The easiest way to
learn about such amplitudes are the simple systems
$^{1,2}$H$-\overline{p}$ and $^{3,4}$He$-\overline{p}$.  Here, we rely
on the absorptive amplitudes extracted from these atoms.  Additional
constraints follow from the chamber experiments~\cite{BIZ74,BAL89}.
Some degree of unification of these data is obtained in terms of
$\overline{N}N$ interaction potentials, in particular the recent
updates of the Paris potential~\cite{PAR05}.  This allows for an
approximate separation of the $S$ and $P$ wave absorption which is
implemented into the optical potential.

The X-ray data from the four lightest atoms indicate the existence of
two quasi-bound $\overline{N}N$ states.  One state occurs in an $S$
wave.  It finds strong support in the $ J/\psi \rightarrow \bar{p}p
\gamma $ decays which allow to pinpoint its quantum
numbers,~\cite{LOI05}. The second quasi-bound state occurs in a $P$
wave,~\cite{LEA05}, and so far has no independent confirmation.  Both
states are reproduced by the Paris model,~\cite{PAR05}, and this
facilitates our discussion of the $ \overline{p}$ atoms.

Second question studied in Section II is the sensitivity of atomic
levels and level widths to the nuclear densities.  A significant
dependence on the input charge densities is found. We attempt a model
independent parametrization of the neutron densities and argue that $
\overline{p}$ atoms test nuclear surfaces but not the single particle
asymptotic density regions.

The optical potential parameters are found via the best fit procedure
to 117 atomic X-ray data.  The fit is improved considerably by the
effect of $\overline{N}N$ $S$ wave quasi-bound state. The other, $P$
wave, state is of no importance but it can explain anomalies observed
in $\overline{p}$ capture on loosely bound nucleons~\cite{WYC01}.

In a number of cases the atomic level widths may be rather precisely
measured for two orbits per atom, the "lower" and the "upper"
one. Such widths are useful to study properties of the nuclear
surface.  Atomic level shifts are less accurate and more difficult to
understand. These difficulties reflect to a large extent the
uncertainties in the understanding of basic $\overline{N}N$
interactions.

In section III, the $R_{rms}$ radii of the neutron density
distributions - $R_{rms}$ - are extracted for several isotopes of Ca,
Zr, Sn, Te and Pb.  This is done on the basis of X-ray data and
radiochemical measurements~\cite{JAS93,LUB94,LUB98,SCH98} which test
the ratios of neutron and proton densities in the region even more
peripheral than that for the X-rays~\cite{WYC96}. The radiochemical
data determine the rate of $\bar{p} n$ capture relative to $\bar{p} p$
capture at very large distances. These are discussed as a separate
issue, since the nuclear region involved in this process is located
about 1 fm farther away from the region tested by the atomic X-rays.
Most of the effort is devoted to the evaluation of uncertainties
involved in this method of density determination.

The appendix  discusses some details of the gradient potential.

The present publication addresses also  three more specific questions:
\begin{itemize}
 \item The optical potential for antiprotons involves uncertain
   quantities: the range and strengths of the $ \bar{N} N $
   interaction and nuclear densities at large distances. Errors due to
   these uncertainties on the atomic level widths are evaluated.
\item The experimental level widths in heavy atoms are determined by
  high moments of the neutron density. However, the nuclear structure
  physics is more interested in the low moments, in particular in the
  rms radius. How well could we determine the latter?
\item Can  antiprotonic data distinguish a neutron skin from a
  neutron halo as defined in Ref.~\cite{TRZ01a}?.
\end{itemize}

A number of  phenomenological optical potentials  have been fitted
to the $ \overline{p}$  X-ray data. Recent results  may be found
in Refs.~\cite{BAT97,FRI05}. The present work is different in
several aspects:
\begin{itemize}
\item  We include  recoil effects in the $P$ wave $ \bar{N} N $ interactions.
\item The constrains from $ Z=1$, $Z=2$ atoms, $ \bar{p} $ absorption
  in flight and $ \bar{N} N $ potential models are accounted for.
\item  The data set is  larger.
\end{itemize}
The constraints allow to obtain the absorptive optical potential
parameters close to those expected from the $ \bar{N} N $ scattering
data. In consequence a good fit to the atomic X-ray data is obtained.

\section {The optical potential }

Atomic energy levels of antiprotons are determined essentially by the
Coulomb and fine structure interactions. In addition, the $
\bar{p}$-nucleus interactions generate level shifts $\epsilon$ and $
\bar{p}$ annihilation generates level broadenings $ \Gamma$. Both
effects may be conveniently described by a complex nuclear optical
potential. The standard potentials, well tested for $\pi$
atoms~\cite{BAT97}, are composed of two terms
\begin{equation}
\label{O1}
 V^{opt} = \Sigma_{p,n} ~~ [ V_S(r) + {\bf \nabla} V_P(r) {\bf
 \nabla}] = V_S + \hat{V}_P,
\end{equation}
with the sum extending over protons and neutrons. Both terms, the
local $V_S$ and the gradient $V_P$ are expected to have a folded form
\begin{equation}
\label{O1S}
 V_{S,P} (r) = \frac{2\pi}{\mu_{\bar{N} N}} a_{S,P} \int d{\bf u}~
 g_{S,P}({\bf u})\rho ({\bf r}-{\bf u}),
\end{equation}
where $\mu_{\bar{N}N}$ is the $ \bar{N} N $ reduced mass and $\rho$ is
the nuclear density. Here $a_{S}$ is expected to resemble the spin
averaged $S$ wave scattering length and $ a_{P}$ the $P$ wave
scattering volume.  Two profile functions $ g_{S,P}$, normalized by $
\int d {\bf u}g_{S,P}({\bf u}) = 1 $, reflect the range of
interactions. The $ \bar{N} N $ annihilation radius $r_o$ is expected
in the range $ 0.8 -1.0 $ fm. Such values follow from phenomenological
or quark models of $\bar{N} N $ interactions. The effective
annihilation radius in models with much shorter annihilation
potentials is similar~\cite{PAR05}. On the other hand, the ranges
involved in Re $V_S$ and Re $V_P$ may be different. In this work, the
same interaction range is assumed for all the components, and it is
left as a free parameter. A Gaussian profile $g$ is used.

The form of $ V^{opt}$ given in eq.(\ref{O1}) is related to the
parametrization of the low energy scattering amplitudes
\begin{equation}
\label{O2}
      f = a_S  + 3  {\bf p} {\bf p}'  a_P
\end{equation}
where ${\bf p},{\bf p'} $ are the relative momenta of the colliding
particles before and after the collision. An important distinction
between $ \bar{N} N$ and $\pi N$ cases is that in the $N\bar{N} $
collisions the nucleon recoil effect is important and the relative
momentum is $ {\bf p}= ( {\bf p}_{N} - {\bf p}_{\bar{N}})/2$. It
involves the nucleon ${\bf p}_{N}$ and antiproton ${\bf p}_{\bar{N}}$
momenta in equal proportions.  As a consequence, gradients over
nucleon wave functions arise in the optical potential. One needs a
formula that generalizes eq.(\ref{O1}). The scattering amplitude
(\ref{O2}) leads now to a new form of the gradient potential
\begin{equation}
\label{O3}
 \hat{V}_G ({\bf r}) = \frac{2\pi}{\mu_{N N}} a_P \frac{3}{4} \Sigma_{\alpha} \int d{\bf r'}
{\varphi}^{*}_{ \alpha}({\bf r'}) (\overleftarrow{\nabla}_{N}
-\overleftarrow{\nabla}_{\bar{N}}) f_P({\bf r}-{\bf r'} )
(\overrightarrow{\nabla}_{\bar{N}} -\overrightarrow{\nabla}_{N})
\varphi_{ \alpha}({\bf r'})
\end{equation}
where $\varphi_{ \alpha}$ are the nucleon wave functions. The
summation over nucleon states $\alpha$ and the integration over
nuclear coordinates ${\bf r'}$ are to be performed. This leads to a
three-term expression for $ V_G $
\begin{equation}
\label{O4}
\hat{V}_G  = \hat{V}_P +   V_N +  V_{mix}
\end{equation}
and
\begin{equation}
\label{O4a}
V^{opt}= \Sigma_{p,n}  ~~ [V_S  + \hat{V}_G] .
\end{equation}
The first term in Eq. (\ref{O4}) corresponds to the standard gradient
potential, $V_N $ is due to the gradients over nucleon functions,
while $ V_{mix} $ follows from mixed nucleon and antiproton
derivatives. In the states studied in experiments the nucleon
dependent part $ V_N + V_{mix}$ contributes about half of the $V_G$
strength, and amounts to a quarter of the total $V^{opt}$. On a
phenomenological level these two terms could be included into the
$V_S$ potential term.  However, the difference arises when one
attempts to relate $a_S $ and $a_P$ to the scattering data. A special
effect comes from the dependence of these terms on the state of the
nucleus, in particular on the angular momentum in the valence
shells. In addition the mixing term may induce some
nucleon-antinucleon correlations in odd-A nuclei. A more detailed
discussion of the gradient terms is given in Appendix A, where some
approximations are also introduced.

\subsection {The choice of  nuclear charge densities }

The nuclear charge densities are well determined in the region of the
nuclear surface between $ c-2a $ and $ c+2a$, where $c$ is the half
density radius and $a$ is the surface diffuseness. These are the
results of electron scattering and muonic atom
experiments~\cite{VRI87,FRI95}. However, the antiprotonic atoms
involve also lower nuclear densities. In Fig.1 are shown the average
radii of $ \bar{p}$ absorption in the "lower" atomic orbits. At those
radii the nuclear charge densities amount to $5\%$ of the central
density. For the "upper" levels these radii are larger by $0.2-0.4$ fm
and involve charge densities smaller by a factor of 2. The absorption
regions are localized in nuclear layers of about 3 fm radial depth. In
these regions, the charge densities are not well known. To indicate
the uncertainty, let us compare the relevant moments of several charge
distributions. Atomic level widths in very high $l$ orbits are given
by expectation values $\Gamma/2 \simeq <n,l| V^{opt}| n,l>$ where $l$
is the angular momentum and $n$ the principal quantum number of an
atomic state.  For large Bohr radii and weak nuclear absorption, the
widths are proportional to $<r^{2l}>$, the $2l-$th moment of the
nuclear density distribution. This reflects the dominant effect of
high centrifugal barriers. In the states actually tested the finer
details of the $ \bar{p}$ wave functions are significant and the
dominant moments are $<r^{2l-2}>$ and $<r^{2l-4}>$. Several relevant
moments of nuclear charge density are compared in Table
\ref{moments}. For a given isotope, all charge density profiles yield
essentially the same rms values. However, the moments of interest may
differ by up to $30\%$ in the highest observable $l$ states. In
particular the many-parameter multi-Gaussian density parametrization
offer much shorter tails. Other extreme cases are given by some
three-parameter Fermi distributions which often generate
unrealistically low (and even negative) densities at very large
distances. To select the "best" charge profile we follow the
"averaging procedure" outlined in Ref.~\cite{LOI01} for hyperonic
atoms.  Thus, the lowest moments $<r^{2}>,<r^{4}>,<r^{6}>$ are
compared for several available charge profiles (Fermi, Gaussian,
multi-Gaussian) and an average density in the sense of average moments
is extracted.  In the cases of Al, S, Ca, Pb, studied in
Ref.~\cite{LOI01} one always finds the average to be the closest to
the profiles given in the most recent compilation by Fricke $et$ $al$
~\cite{FRI95}. The same is found in all Ni, Zr and Sn isotopes studied
here. We use the parameters of Ref.~\cite{FRI95} as the basis of our
calculations.  The only exception is $^{16}$O, where the sum of
Gaussians and the deformed nuclei U, Th, where monopole density
component were used~\cite{VRI87}.  Later, in specific cases, the
comparison with other densities is presented.

\begin{figure}[ht]
\includegraphics[width=0.5\textwidth]{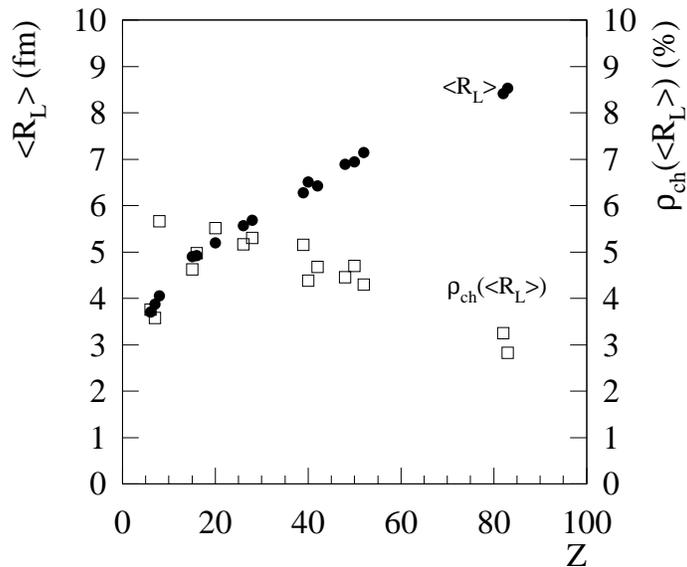}
\caption{Dots - the average antiproton absorption radii $<R_L> $
  (weighted by $ \psi_{atom}^2$ Im $V^{opt}r^2$ ) calculated for the
  "lower" atomic level, left scale in fm.  Squares - the charge
  densities at these points, given on the right scale as $\%$ of the
  central charge densities. The bottom scale - atomic numbers Z.
  Calculations are done with densities from Ref.~\cite{FRI95}.}
  \label{fig:radius}
\end{figure}

\begin{table}[ht]
\caption{Moments $<r^{2l}>= \int d\bf{r} \rho(r) r^{2l} / \int d\bf{r}
  \rho(r)$ of the charge density profiles in $ 10^{2l-3}$ fm$^{2l}$
  units. The densities based on $\mu $ atom data -first column,
  electron scattering data - other columns: 2pF - two parameter Fermi,
  HO - harmonic oscillator, SOG - multigaussian.  }
\label{moments}
\begin{center}
\begin{tabular}{|c|c|c|c|c|c|} \hline
          Atom & moment  & 2pF($\mu$)  \cite{FRI95} & 2pF(e) \cite{VRI87} &  HO(e) \cite{VRI87}        &  SOG(e) \cite{VRI87}  \\ \hline
  $^{16}$O     &rms      & 2.72               & 2.72             &   2.72    & 2.71  \\
               & $<r^4>$     & 9.0               & 8.7               &   8.2      & 8.4  \\
               & $<r^6>$   & 1.6               & 1.5              &   1.2      & 1.3  \\ \hline

  $^{32}$S     & rms& 3.28               & 3.25               &           & 3.26      \\
               & $<r^4> $   & 17.2           & 16.9              &           & 1.69       \\
               & $<r^6>$   & 3.75           & 3.81              &           & 3.82  \\   \hline

  $^{58}$Ni    & rms     & 3.79           & 3.78             &           & 3.77       \\
               &  $<r^6>$  & 7.5            & 7.6               &          & 6.7   \\
               &  $<r^8>$  & 2.6           & 2.5               &           & 1.9   \\ \hline

  $^{116}$Sn   &rms       & 4.64               & 4.65             &          & 4.63   \\
               & $<r^8> $   & 8.6                & 9.8              &          & 7.9    \\
               & $<r^{10}>$   & 4.2                & 5.1              &         & 3.7   \\
               & $<r^{12}>$   & 2.5                & 3.3              &         & 1.7   \\  \hline

  $^{124}$Sn   &rms      & 4.68               & 4.66             &          & 4.69    \\
               & $<r^8>$   & 9.3                & 9.5              &          & 8.4      \\
               & $<r^{10}>$  & 4.6                & 4.8              &         & 3.7    \\
               & $<r^{12}>$  & 2.6                & 2.9              &         & 1.8   \\  \hline

  $^{208}$Pb   &rms       & 5.504              & 5.520            &          & 5.503   \\
               & $<r^{12}>$  & 11      & 13             &          & 9.3      \\
               & $<r^{14}>$  & 8.7      & 11.2            &         &
  6.2    \\ \hline
\end{tabular}
\end{center}
\end{table}

\subsection {The  parametrization of nuclear densities }

To understand the $ \bar{p}$ atomic data, one needs a reliable
extrapolation of the densities to very large distances. The related
problems are visualized in Fig.2. For two Ca isotopes the densities
were calculated with a HFB-SkP model~\cite{SMOL}, and fitted at large
distances by a two-parameter Fermi profile
\begin{equation}
\label{10a} \rho(r) =1/(1+ \exp[(r-c)/a(r)].
\end{equation}
Instead of a constant diffuseness one has to introduce certain
functions $a(r)$, which are plotted in Fig. 2.  The dependence on
radial distance is rather distinct. There are clear advantages of the
2pF profile, but with a constant $a$ it does not reproduce the density
in far away regions.  That result is also fairly model independent and
the shapes of $a(r)$ in Fig. 2 indicate a certain degree of
universality.  At distances $ c+3a \leq r \leq c+6a $ one finds
essentially the same slope of $a(r)$ for a wide range of nuclei in
Hartree-Fock, Hartree-Fock-Bogolubov~\cite{WYC96,SMOL} and
Relativistic Mean Field calculations~\cite{LAL05}. However, at radii
beyond this region, different nuclear models may produce different
behavior.

The average annihilation radii $r_a$ in Ca are marked by arrows. In
the lower and upper states one has $ r_a = 4.9 $ fm and $5.3$ fm
respectively and $a(r)$ is seen to be fairly stable at these
ranges. However, the radiochemical data involve larger radii from
$\approx 5.1$ fm up to $ r \approx 7.5 $ fm~\cite{WYC96}.  The
diffuseness parameter $a(r)$ is seen to fall down in this region. The
difficulty involved in the radial dependence of $a(r)$ is moderated by
the optical potential which involves folded densities.  With the
folding range of $\approx 1 $ fm the corresponding downfall of
$a_{folded}(r)$ is pushed away by $\approx 0.5 $ fm.

\begin{figure}[ht]
\includegraphics[width=0.6\textwidth]{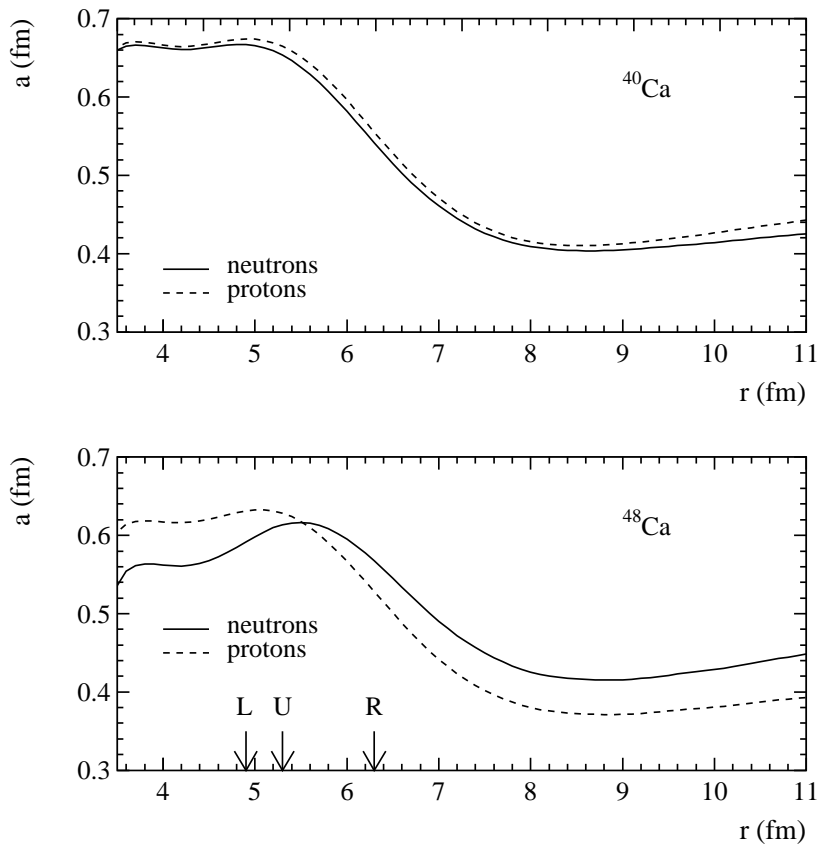}
\caption{The proton and neutron densities in two Ca isotopes
  calculated in a HFB shell model have been approximated by the two
  parameter Fermi distribution with constant $c$ and variable
  $a(r)$. Functions $a(r)$ are plotted to show it's asymptotic
  behavior at large distances. The arrows indicate absorption radii
  for the lower (L), upper (U) and radiochemical (R) experiments,
  correspondingly.} \vspace{.0cm} \label{fig:ca}
\end{figure}

In order to understand $a(r)$ the first question to answer is where
the single particle asymptotic limit is reached and what is its form.
For a single neutron of angular momentum $L$ bound in an external
potential the asymptotic wave function is described by
\begin{equation}
\label{11}
\varphi_{\alpha}\sim h_L( irk_{\alpha})= \frac{\exp(-rk_{\alpha})}{r}
W_L(\frac{1}{k_{\alpha}r})
\end{equation}
where $ k_{\alpha} = \sqrt{2m_N E_{\alpha}}$, $E_{\alpha}$ is the
neutron binding energy and $W_L$ is a polynomial given by the Hankel
function of order $L$ (see e.g.~\cite{BER82}). This wave function
allows to scale the asymptotic density $\rho(r) =
\varphi_{\alpha}(r)^2$. The ratio of two densities $\rho(r)/\rho(r_o)
$, taken at two points $r_o$ and $r$, allows to find
\begin{equation}
\label{12}
\frac{1}{a(r)} = [ \frac{1}{a(r_o)}- 2 k_{\alpha} + 2\ln(
\frac{W_L(r_o)r }{ W_L(r)r_o }) ] \frac{r_o}{r} +2 k_{\alpha} .
\end{equation}
This formula reproduces the shapes of $a(r)$ in the region of $ r
\approx 5-8 $ fm with $k_{\alpha} \approx 1.25 $ fm$^{-1}$ which
corresponds to $E_{\alpha}\approx 30$MeV. On the other hand, the
separation energies are respectively 16 MeV ($^{40}$Ca) and 10 MeV
($^{48}$Ca). Thus the "true" single particle asymptotic density given
by valence neutrons begins farther away. The second observation is
that the effective $k_{\alpha}$, which reproduces Fig. 2, corresponds
to bindings larger than the average binding weighted by contributions
of neutron orbitals to the total neutron density. The behavior of a
nucleon at large distances involves virtual excitations of the
residual system and a sizable fraction of $k_{\alpha}$ is due to the
nuclear correlations. Contrary to a frequently expressed belief, the
radii tested by antiprotons do not represent the far away distances of
single particle asymptotic wave functions, and in this sense are more
interesting for the nuclear structure research.

In practical terms the shortcomings of constant $a$ can be corrected
by the $a(r)$ given by eq. (\ref{12}) with $k_{\alpha} \approx 1.25 $
fm$^{-1}$. Because of the large value of $k_{\alpha}$ the result is in
practice independent of the value of $L$.  Relation (\ref{12}) is used
here to interpolate between two values of the diffuseness
parameter. The initial value $a(r_o)$ taken at $r_o =c+3a $ is due to
the full complexity of the nuclear structure which includes the
average field and effects of nuclear correlations. It is kept as a
free parameter, $\it{the}$ $a $, to be determined from experiments and
reproduced by models. Such a procedure serves only as a guiding
principle for the best correlation of the atomic X-ray data and the
radiochemical data. With an improved optical potential and more
precise data it should be repeated with specific nuclear models.
However, at this stage the proper choice of $a(r_o)$ and the form of
the proton (charge) densities are more urgent questions.

\subsection {Constraints on the optical potential parameters }

The $N\bar{N} $ amplitudes of eq.(\ref{O2}) are related to the
amplitudes tested in scattering experiments.  The latter extrapolated
to the $N\bar{N} $ threshold yield scattering lengths and scattering
volumes which parametrize the low energy scattering. The relation of
the experimental lengths and volumes to those required in the optical
potential is not direct: first - $a_S(E)$ and $a_P(E)$ are strongly
energy dependent and one needs these amplitudes for bound particles
and second - some nuclear many body corrections may arise. An
additional difficulty is related to the large number of $N\bar{N} $
partial waves involved, and at this stage of research one can operate
only with the spin averaged values.  So far, the safest method to find
the optical potential was to extract the potential parameters from the
best fit to the atomic data~\cite{BAT97,FRI05}. In our work we adopt a
mixed procedure: the parameters are semi-free, subject to constraints
from other experiments. In addition, we are guided by the Paris model
of $N\bar{N} $ interactions.

At the nuclear surface $ \bar{p} $ may interact with quasi-free but
bound nucleons.  In the $N\bar{p}$ system the relevant energy is
negative since both particles are bound and some recoil energy is
taken away by the relative motion of the $N\bar{p} $ pair with respect
to the residual $A-1$ nucleons.  Hence, one needs to know the
amplitude $f$ below the $N\bar{p}$ threshold
\begin{equation}
\label{PO2}
      f = a_S(-E_{B}-E_{rec})   + 3   {\bf p} {\bf p}' a_P(-E_B-E_{rec})
\end{equation}
where $E_B$ is the sum of antiproton and nucleon separation energies
and $E_{rec}= p_{rec}^2/2 \mu_{rec}$ is the recoil energy. The
distribution of the recoil momenta - $ p_{rec}$ - is calculable from
the Fourier transforms of the antiproton and nucleon wave functions $
\Psi_{\bar{p}}({\bf r}) \varphi_{N}({\bf r})$. In practice, such
calculations can be done easily in the lightest atoms: deuterium and
helium~\cite{WGN85}. In the "lower" ($l=0$) and "upper" ($l=1$)
orbitals of $^2$H $\bar{p}$ atoms one obtains average values
$<E_{rec}>= 9$ and $ 4 $ MeV respectively. In heavier nuclei such
calculations are less reliable but the spectrum of $p_{rec}$ was
measured with $\bar{p}$ stopped in Ne streamer
chamber~\cite{BAL89ne}. The $p_{rec}$ distribution peaks at $\approx
180 $ MeV and gives $<E_{rec}>\approx 11 $ MeV. This value is used in
further calculations. In addition, we estimate a 5 MeV difference of
$<E_{rec}>$ in the "lower" and the "upper" atomic state.

\begin{figure}[ht]
\includegraphics[width=0.75\textwidth]{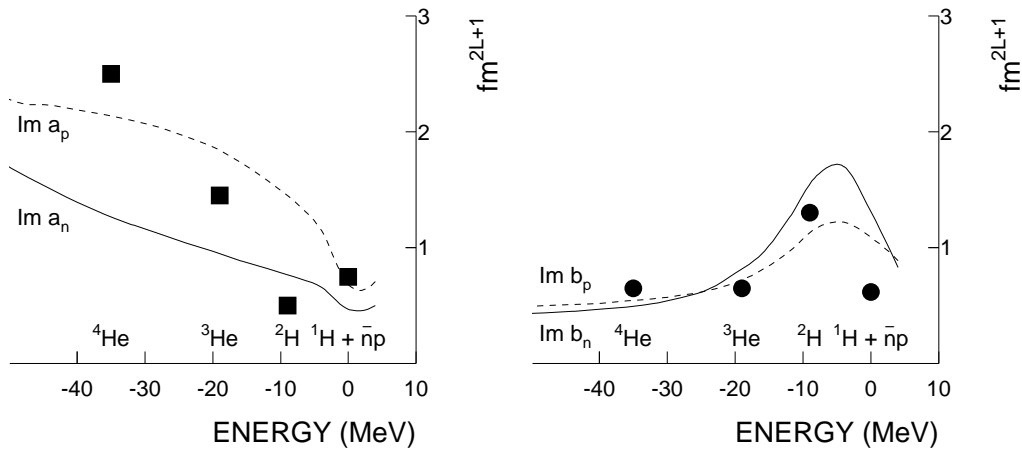}
\caption{Left panel: the absorptive parts of $\bar{p}N $ $S$-wave
  amplitudes in the sub-threshold region extracted from the atomic
  level shifts and widths in $^1$H, $^2$H, $^3$He, $^4$He.  Im
  $a_S(E)$ averaged over spins and isospin of the nucleons are given
  by squares located at the corresponding values of $E_B+<E_{rec}>.  $
  The curves are calculated with the recent updates of Paris $N\bar{N}
  $ potential,~\cite{PAR05}.  The dotted lines refer to $p \bar{p} $
  and the continuous lines to $ n \bar{p} $ systems.  Right panel: the
  absorptive parts of $\bar{p}N $ $P$-wave amplitudes in the
  sub-threshold region extracted from the atomic level shifts and
  widths in $^1$H, $^2$H, $^3$He, $^4$He (circles) The curves -- Paris
  potential calculations.}
\label{imas}
\end{figure}

The energy dependence of the $N\bar{p}$ scattering amplitudes in the
sub-threshold region may be obtained, to some extent, from the $^1$H,
$^2$H, $^3$He, $^4$He atoms since the nucleon separation energies in
these nuclei span the region from 0 to 21 MeV.  Calculations based on
the multiple scattering series summation from Ref.~\cite{WGN85} and
data from Refs.~\cite{AUG99,AUG99a,GOT99,SCH91}, are presented in
Fig.~(\ref{imas}).  From the level shifts and widths in these elements
one can extract averaged absorptive parts of Im $a_S$ and Im $a_P$ via
a best fit procedure~\cite{LEA05}. The nucleon binding energies
characteristic for surfaces of large nuclei locate the $E_B+E_{rec}$
energies in the sector (40 to 15) MeV.  Corresponding values of Im
$a_{S}$ and Im $a_{P}$ indicate the strengths of the absorptive
optical potentials expected in nuclei. Two results are of significance
in the analysis of the optical potential. First, the $S$-wave
absorption strength Im $a_S$ increases with the decreasing energy. The
physics behind it is related to a broad quasi-bound $N\bar{N}$ state
which is indicated by the atomic data and generated by the Paris
potential. Its existence is also inferred from the $\bar{p}p $
correlations observed in $ J/\psi \rightarrow \bar{p}p \gamma $
decays,~\cite{LOI05}. The impact of such phenomenon is discussed in
the next section, it is clearly seen in the comparison of "lower" and
"upper" widths.  Second, a resonant-like behavior arises in a $P$-wave
close to -10 MeV. Figure (\ref{imas}) indicates that a similar effect
is also generated by the Paris potential model, where it is attributed
to the iso-triplet, spin-singlet $^{31}P_1$ quasi-bound state.  It
affects antiproton capture on very loosely bound nucleons
\cite{WYC01}.

A different constraint on the isospin structure of Im $a_S$ follows
from studies of $\pi$ mesons produced by $\bar{p}$ stopped in
chambers~\cite{BIZ74,BAL89}. These experiments yield ratios of
annihilation rates $ R_{n/p} =\sigma ( \bar{p}n)/ \sigma ( \bar{p}p)$
given in Table \ref{chamber}. Since the $\bar{p}$ are stopped in light
elements the interaction involves $S$ waves and the cross sections are
related to the absorptive amplitudes $\sigma ( N\bar{p})\sim {\rm Im}
\: a_S(N\bar{p})$ for the $n\bar{p}$ and $p\bar{p}$ pairs,
respectively. Inspection of Fig. (\ref{imas}) shows that the $
R_{n/p}(S)$ for the $S$ waves is well reproduced by the Paris
potential model. The same model predicts $ R_{n/p}(P)\approx 1$ for
the average of $P$ waves at energies of our main interest. These
constraints are considered as possible improvements of our optical
potential. It turns out that the condition $ R_{n/p}(P)\approx 1$ is
consistent with the atomic data.  In this respect our results follow
the findings of Ref.~\cite{FRI05}.  The chamber result $
R_{n/p}(S)\approx 0.5$ leads to no or a marginal improvement over that
for $ R_{n/p}(S)\approx 1$.  On the other hand, it changes the neutron
density radii extracted from $\bar{p}$ atoms.

\begin{table}[ht]
\caption{ The  experimental antiproton capture ratios $R_{n/p}=
\sigma (\bar{p}n)/ \sigma ( \bar{p}p)$ extracted  from capture in
flight.}
\label{chamber}
\begin{tabular}{|lc|c|}
\hline
Element &  $R_{n/p}$    &  reference    \\
\hline
$^{2}$H   &  0.81(3)    & \cite{BIZ74}    \\
$^{3}$He  &  0.47(4)    & \cite{BAL89}    \\
$^{4}$He  &  0.48(3)    & \cite{BAL89}    \\ \hline
\end{tabular}
\end{table}

\subsection {The constrained  best-fit potential }

An overall best fit to the 117 atomic data was performed. It includes
measurements in N, C, O, P, S, Cl, Ca, Fe, Co, Ni, Y, Zr, Cd, Mo, Sn,
Te, Pb, Bi, Th and U, in total 37 isotopes. Several precise results
were excluded: $^{6,7}$Li - as the optical potential may not be well
applicable there, $^{18}$O - due to the large uncertainty of the
charge density profile as exemplified already in $^{16}$O, Yb - due to
its uncommon deformation. An anomalous lower width in $^{58}$Fe was
dropped and several very old and uncertain results have not been
considered.

One cannot find the best fit parameters jointly with the details of
uncertain neutron densities in each individual isotope. To obtain an
overall best fit the following strategy was adopted. An overall trend
of the neutron rms radius as a function of the neutron excess $ \delta
= ( N-Z)/(N+Z)$ was assumed and for $Z > 10$ a linear interpolation
was tried. Two such trends were fitted before to the atomic $ \bar{p}$
data. One, called "hadronic" is indicated by the hadronic scattering
data~\cite{JAS04}
\begin{equation}
\label{OP1} \Delta R_{np}  \equiv   R_{rms}(n)-R_{rms}(p) = -
0.09(2) + 1.46(12)~ \delta.
\end{equation}
It is close to the trend  obtained  by Friedman $et$ $ al$ as a
result of  pionic atom data  and the best-fit to the antiproton
X-ray data~\cite{FRI05}.
 Another "atomic"  trend follows from  the $\bar{p}$ atomic data
 analyzed with   zero
$\bar{N} N $ annihilation range  and    $R_{n/p}=1$,~\cite{TRZ04a},
\begin{equation}
\label{OP2}
R_{rms}(n)-R_{rms}(p) = -0.03(2) + 0.90(15)\delta
\end{equation}
Both slopes can be supported by nuclear model calculations and it is
appropriate at this point to remind that nuclear models are not able
to predict the neutron radii and we have to fit some parameters to
experimental data.  This point is particularly strongly stressed by
Furnstahl~\cite{FUR02} and a similar point of view is taken in the
recent extensive calculation by Ring $et$ $ al$.~\cite{LAL05}. Neither
of these trends indicated above needs to be true, some nuclei lie far
away from the averages given above. Here, we use the slope to get an
initial insight into the best fit possibilities. Next with the best
parameters for the optical potential, we find the best results for the
neutron excess in some nuclei. This is done, following the procedure
of Ref.~\cite{TRZ01a}, in terms of free neutron diffuseness $ a_n$
(neutron halo, $ c_n= c_p$) or neutron half-density parameter $ c_n$
(neutron skin, $ a_n = a_p$). An additional question is whether the
inclusion of extra data, the radiochemical measurements or other
experiments, determines some correlation of these two parameters.

\begin{table}[ht]
\caption{Overall parameters for the optical potential fitted to atomic
  data: 78 - lower level shifts and widths, 39 - upper level
  widths. In the first line the slope follows eq.\ref{OP2}. With other
  "hadronic" cases the slope was varied but close to the one given by
  eq.\ref{OP1}. The best result is obtained with $\Delta R_{np}^h = -
  0.10(2) + 1.65(5)\delta $. The $r_{rms}$ denotes the root mean
  square radius of the folding Gaussian profile. All charge densities
  come from Ref.~\cite{FRI95}. Factor $f_i = 4/3 $ for protons and 2/3
  for neutrons describes the chamber result, $a_n$ or $c_n$ denotes
  the neutron density free parameter.  }
\label{potential}
\begin{center}
\begin{tabular}{|l|c|c|c|c|c|c|c|}
\hline
 "slope"& parameter & data  & $ \chi^2 $ & $ \chi^2/N $ &  $ a_S$ [fm]  & $ a_P$ [fm$^3$] & $r_{rms} $[fm] \\
\hline
Eq.\ref{OP2}      &  $ a_n$ &  all    &  293 & 2.50  &  -1.70 - 0.95i      & -.01 - 0.39i    & 0.80      \\
$\Delta R_{np}^h$ &  $ c_n$ & all     &  285 & 2.43  & -1.70 - 0.95i       & -0.0 - 0.35i     & 0.79    \\
$\Delta R_{np}^h$ & $ a_n$& all     &  262 & 2.24  &  -1.65 - 0.85i     & -0.0 - 0.41i     & 0.79    \\
$\Delta R_{np}^h$ & $ a_n$& all   &  262 & 2.24  &  -1.60 -$f_i$0.95i & -0.0 - 0.39i     & 0.79    \\
$\Delta R_{np}^h$ & $ a_n$& lower &  160 & 2.24  &  -1.60 - 0.95i       & -0.0 - 0.41i     & 0.79    \\
$\Delta R_{np}^h$ & $ a_n$& upper &   87 & 2.05  &  -1.60 - 0.62i      & -0.0 - 0.41i     & 0.79    \\
 \hline
\end{tabular}
\end{center}
\end{table}

The initial best fit results are given in Table \ref{potential}. The
faster $\Delta R_{rms} $ slope deduced from the hadronic scattering
offers a better $ \chi^2$. This result differs from the result
obtained in ref.,~\cite{TRZ04a}, as it is based on both widths and
level shifts while ref.  ,~\cite{TRZ04a}, used level widths. The best
result $\Delta R_{np} $ = $ - 0.10(2) + 1.65(5)\delta $ obtained with
a change of the diffuseness parameter $a_n$ corresponds to line 4 in
Table III. The difference of our result and that given by
eq.(\ref{OP1}) and Ref.~\cite{FRI05} is due mainly to the input charge
densities.  Here, the more recent muon-based data~\cite{FRI95} are
used while results of Ref.~\cite{FRI05} are based mainly on the
electron data. As discussed in previous sections and indicated in
Table I, the muonic data are close to the "average" results and in
this sense seem to be preferable. In addition,

Another minimum of $ \chi^2$ is obtained with changes of the
half-density radius $ c_n$ at a very high slope $\Delta R_{rms} $ = $
- 0.10 + 2.0 ~\delta $. Although the total $ \chi^2 $ is much worse in
this case, it turns out to be better for some specific isotopes. This
question is discussed in the next section. Let us also notice that the
best potential parameters are fairly close for both the "atomic" and
the "hadronic" slopes.

The preference found in chamber experiments~\cite{BIZ74,BAL89} for a
stronger $S$ wave annihilation on protons, $R_{n/p}(S) \approx 0.5 $,
is now introduced into the optical potential parameters. The effect is
given in line 4 of Table \ref{potential}. It does not change the best
fit, but is accepted by the data. It is also weakly reflected in the
analysis of radiochemical data as the absorption in $P$ waves is
dominant.  The effective ratio of the absorption rates $R_{n/p}$
depends on the partial wave in the antiproton-nucleon system. For each
atomic state the mixture of $S$ and $P$ wave is slightly different. We
calculate a kind of average value characteristic for all $Z,n,L$
states considered in this analysis. For the best fit potential of
table \ref{bestpotential} one obtains $R_{n/p}= 0.86(4)$ as the best
average representation for the capture ratio $\sigma (\bar{p}n)/
\sigma ( \bar{p}p)$ at large distances tested in the radiochemical
experiments.  The uncertainty given in parentheses describes the
dispersion of $R_{n/p}$ obtained in this way.  A more detailed
discussion of this point may be found in Ref.~\cite{RIK05}.

Lines 5 and 6 in Table \ref{potential} give separate best fits to the
lower and upper levels. To find the best solution on those limited
data sets only Im $a_S$ was varied. The motivation for such a choice
comes from the shape Im $a_S(E)$ in Fig.(\ref{imas}). The best fit
absorptive parts compare well with the values of Im $a_S(E)$ at the
characteristic subthreshold energies of about -30 MeV.

\begin{table}[ht]
\caption{The best fit potential based on X-ray data, consistent with
 the chamber experiments, the lightest atoms and the $N-\bar{N} $
 Paris potential.  $f_i = 4/3 $ for protons and 2/3 for
 neutrons. $a_S("upper")$ is to be used for the upper levels and
 $a_S("lower")$ for the lower ones.  }
\label{bestpotential}
\begin{center}
\begin{tabular}{|c|c|c|c|c|c|}
\hline
$ \chi^2 $&  $ \chi^2 /N$        &  $a_S("upper")$[fm]  & $ a_S("lower")$[fm] &    $ a_P$ [fm$^3$]  &    $ r_o$[fm] \\
\hline
 247   &  2.11&   -1.60 -i 0.74$f_i$ & -1.60 - i 1.10$f_i$   & -0.0 - i0.39     & 0.79    \\  \hline
\end{tabular}
\end{center}
\end{table}

Certain incompatibility in the description of lower and upper widths
was noticed already in Ref.~\cite{BAT97}. This effect is also
reproduced here and it is indicated in Table \ref{potential}. The
upper levels require weaker absorption and the explanation comes from
the left panel of Fig.(\ref{imas}). The $\bar{p} $ in the upper level
encounters less-bound nucleons and the recoil energy is also smaller.
As discussed above, one expects the difference in recoil of some 5 MeV
and the difference in the average bindings of some 3 MeV. The central
energy involved in $a_{S,P}(-E_{B}-E_{rec}) $ amounts to about - 35
MeV in the lower levels and about - 27 MeV in the upper levels. The
total 8 MeV energy shift in the argument of Im $a_S(E)$ may reduce its
value by $0.4$ fm as required by the X-ray data.  Table
\ref{potential} indicates that such a change reduces the total $
\chi^2 $ from 262 to 247. The final results are summarized in Table
\ref{bestpotential}. This potential does not offer the best fit to the
data, however it offers the best fit under the chamber constraint on
the S wave absorption $ R_{n/p}(S) = 0.5 $.  Relaxing this condition
one can improve the total $ \chi^2 $ by 2-3 units. In addition, it is
possible to improve the fit to the data by relaxing the other
condition $ R_{n/p}(P) = 1 $. There is no direct experimental
indication for this P wave constraint, we are motivated entirely by
the Paris potential calculations.

\section {The neutron radii}

The differences of neutron and proton mean square radii, extracted
from from several atoms, are given in Tables \ref{Sn} and
\ref{Pb}. These results indicate two basic problems:

First, the extracted $R_{rms}(n)-R_{rms}(p)$ depend rather strongly on
the charge density input. In particular, the results in the Sn
isotopes depend on the $<r^{12}>$ moments dominating the upper level
width.  For two charge profiles in $ ^{112,116,120,124}$Sn : one from
Ref.~\cite{VRI87}(electron scattering) and the other from
Ref.~\cite{FRI95}( $\mu$ atoms) the ratios $<r^{12}>_{e}$ /
$<r^{12}>_{\mu}$ are :$ ~ 1.63, ~ 1.32,~ 1.42 ,~ 1.09$,
respectively. These differences reflect on the differences in neutron
radii extracted from the antiproton data.  As discussed above, the
charge densities given by Fricke $et$ $ al$~\cite{FRI95} are close to
"average" densities generated in several $\mu$ atom and electron
scattering experiments. In this sense these results are more likely
than the others.

Second, the minimum $ \chi^2 $ in each isotope was obtained either by
enhancing the diffuseness parameter $a_n$ or half density radius
$c_n$.  In most cases both minima offer good $ \chi^2_{pdf} \approx 1
$ and additional data are required to determine the nuclear surface
shape. However, in $^{112}$Sn, $^{116}$Sn, $^{90}$Zr and $^{208}$Pb
the fit is bad and $ \chi^2_{pdf} \approx 3$.  The discrepancy comes
from the level shifts.  In these cases the change of the half-density
radius offers better $ \chi^2 $ and such case is given in Table
\ref{Pb} for Pb.  The neutron radii obtained via the $c_n$ extension
are excessively large and run into conflict with most of the other
data.  We exemplify this situation in the case of Sn and Pb atoms, but
it is typical to other large nuclei. In the Pb nucleus the two minima
for $R_{rms}(n)-R_{rms}(p)$ yield very different results. The solution
obtained be enlarging $a_n$ is close to the results obtained in (p,p')
scattering experiments~\cite{STA94}. Presumably it is the one that is
physically acceptable. The uncertainty related to the charge density
input is rather large but again there are good reasons to favor the
last column based on the Ref.~\cite{FRI95}.  It is interesting to note
that the second solution characterized by the extension of $c_n$ and
large difference $R_{rms}(n)-R_{rms}(p)$ of 0.5 fm is close to the
result first obtained in the neutron/proton pickup experiment of
Koerner and Schiffer~\cite{KOR71}. This solution, if it represents the
reality, has an interesting astrophysical significance~\cite{FUR02}.

The same interpretation follows from the results obtained in Sn.  The
neutron radii obtained with enhanced $ a_n$ are close to those
obtained in the proton scattering experiment~\cite{RAY79} that yields
$R_{rms}(n)-R_{rms}(p)=0. 25(5)$ fm in $^{124}Sn$, larger than 0.18(7)
extracted via the dipole-state excitation method~\cite{KRA99}. The
enhancement of $ c_n$, allowed by the $\chi^2$, yields very large
neutron radii not confirmed in other experiments.

It was argued in Ref.~\cite{TRZ01a} that the additional data needed
to pinpoint the nuclear surface shape come from the radiochemical
measurements. With an optical potential derived from zero range $
\bar{N} N $ interactions, these data favored the halo type
solution. Such conclusion is also supported here, and exemplified in
Table \ref{Sn} by the $^{124}$Sn case. It also favors the halo type
solution but the extracted $R_{rms}(n)-R_{rms}(p)$ are larger and
closer to the hadron scattering results.  The same effect occurs in
the case of Pb nucleus. The favored neutron radius excess
$R_{rms}(n)-R_{rms}(p)= 0.22(3) $fm given in Table \ref{Pb} is $0.05$
fm larger than the radius obtained with different optical potentials
in Ref.~\cite{KLO06}. There are two factors contributing to that: the
first and the dominant one is the chamber data that enforces smaller
$R_{n/p}$, and second the $a(r)$ given by nuclear models tends to be
smaller at large distances. Due to both factors the neutron radius
obtained from the X-ray data is slightly smaller than the radius
obtained from the joint X-ray and radiochemical data.

\begin{table}[ht]
\caption{The $R_{rms}(n)-R_{rms}(p)$, in fm units, extracted from the
 X-ray data in antiprotonic Sn, Te, Zr and Ca atoms. The first column
 refers to charge density profile the second indicates the free
 parameter of the neutron density. As discussed in the text, the
 results obtained with charge density from Ref.~\cite{FRI95} are the
 favored ones. The last column and the entries marked by $^*$ are
 extracted from the X-ray and radiochemical data. } \label{Sn}
\begin{center}
\begin{tabular}{|l|l|cccc|c|} \hline
$\rho_{charge}$&  & $^{112}$Sn   &$^{116}$Sn &$^{120}$Sn
&$^{124}$Sn  &$^{124}$Sn$^*$  \\ \hline
\cite{FRI95}   & $a_n$   & 0.21(5)      & 0.22(4)   &  0.22(4)   & 0.23(4)    & 0.26(4)                \\
2pF\cite{VRI87}& $a_n$   & 0.12(4)      & 0.15(4)   &  0.16(4)   & 0.21(4)    & 0.25(3)    \\
2pF\cite{VRI87}& $c_n$   & 0.25(5)      & 0.31(6)   &  0.35(8)   & 0.42(7)
   & 0.45(6)    \\ \hline \hline
               &                & $^{122}$Te   &$^{124}$Te &$^{126}$Te  &$^{128}$Te  &  $^{128}$Te$^*$             \\ \hline
\cite{SHE89}   & $a_n$   &  0.10(7)     &  0.04(4)  & 0.12(5)    &$0.08^{+.07}_{-.04}$& 0.15(5)    \\  \hline
\hline
      &   &$^{130}$Te &$^{130}$Te$^*$&$^{90}$Zr
&$^{96}$Zr   &$^{96}$Zr$^*$ \\ \hline

\cite{SHE89},\cite{FRI95}   & $ a_n$   & 0.10(5)     &  0.17(4)  &
$0.08^{+.01}_{-.04}$&  0.12(3)  &0.16$^{+0.06}_{-0.08}$     \\
\hline \hline
               &                &$^{40}$Ca&$^{42}$Ca  &$^{44}$Ca  & $^{48}$Ca & $^{48}$Ca$^*$ \\ \hline
 \cite{FRI95}& $ a_n$ & -0.09(9)&  0.01(6)     &.02(6)        & $0.09^{+.06}_{-.08}$& $0.10(7)$           \\ \hline

\end{tabular}
\end{center}
\end{table}

\begin{table}[ht]
\caption{ The $R_{rms}(n)-R_{rms}(p)$ differences extracted from
  antiprotonic $^{208}$ Pb atoms. The entries in parentheses [ ] give
  the corresponding $\chi^2$ given mostly by the level shifts.  Second
  line gives the difference of neutron and proton diffuseness
  parameters $ a_{np} = a_n - a_p $} \label{Pb}
\begin{center}
\begin{tabular}{|l|c|c|}
\hline
 parameter &  $\rho_{charge}$  \cite{VRI87}  &  $\rho_{charge}$  \cite{FRI95}  \\
\hline
$  a_n  $ & 0.16(3) [10]   &  0.22(3)[10]    \\
 $  a_{np}$ & 0.11         &  0.15           \\
$  c_n  $ & 0.47(8) [3]    &  0.55(8)[3]     \\ \hline
\end{tabular}
\end{center}
\end{table}

\section {Conclusions }

In this paper a step was taken to describe the antiprotonic atom data
in a semi-phenomenological way. The best-fit optical potential was
constrained by the atomic data from the $\bar{p} $ hydrogen, deuterium
and helium atoms. Additional information on the $S$ wave isospin
structure was extracted from the chamber low-energy $\bar{p}$
data. All these constraints refer to the absorptive
potentials. Following results have been obtained:

\begin{itemize}
\item A consistent description of the annihilation parameters in
  antiprotonic X-ray data, chamber data and the updated Paris
  potential model, which incorporates the $\bar{N}N$ scattering.
\item This consistency allows to separate the $S$ and $P$-wave
  contributions to the optical potential for antiprotons and estimate
  fairly precisely the ratio of $\bar{p}n$ and $\bar{p}p$ annihilation
  at very distant nuclear surfaces. The corresponding parameter
  $R_{n/p}= 0.86(4) $ averaged over a range of nuclei is obtained at
  these distances and it allows to discuss jointly the X-ray data and
  the radiochemical data.
\item A definite energy dependence of the $\bar{N}N$ scattering
  amplitudes in the subthreshold region is indicated by the lightest
  atom data, chamber data and the $\bar{N}N$ Paris potential. For the
  $S$ wave it is reflected in the widths of upper and lower levels in
  heavy atoms. For $P$ waves there is an indication of a fairly narrow
  $\bar{N}N$ quasi-bound state. It is likely that such a state has
  sizable effect in a small sector of the radiochemical data taken on
  loosely bound protons. On the other hand, apart from the deuteron,
  the X-ray data offer no convincing evidence for such a state.  New
  X-ray experiments performed with nuclei of small nucleon separation
  energies would be helpful to resolve this question.
\item The $\bar{p}$ X-ray data, although related to the nuclear
  surface, may supply important information on the neutron radii.
  However, these data taken by itself cannot tell precisely if the
  neutron excess forms an extended half-density radius, enlarged
  diffuseness or some specific correlation of both parameters.
  Jointly with the scattering and/or radiochemical data the $\bar{p}$
  results favor the neutron profiles of enlarged diffuseness.
\item One important source of uncertainty in the description of the
  $\bar{p}$ nucleus interactions is the real part of the optical
  potential and the related level shifts. It is complicated by the
  increasing evidence of $\bar{p}N$ quasi-bound states. Precise
  measurements of the atomic fine structure, in particular in very
  light atoms, would be very helpful to resolve this question.
\end{itemize}

\appendix
\section {The gradient potential }

The P wave collisions of $ N$ and $ {\bar{N}}$, described in their
center of mass system, involve the relative momentum $ {\bf p}= ( {\bf
p_{N}}- {\bf p_{\bar{N}})/2}$. It generates the gradient potential
given by expression (\ref{O3}) in the main text which contains the
nucleon wave functions and their gradients. These are not easily
reducible to the nuclear densities. Summation over nucleon states and
calculations of the derivatives in eq.(\ref{O3}) yield involved
expressions for $ V_G $. The input uncertainties and experimental
errors call for a simpler result.  To obtain it let us remind that in
the "upper" high angular momentum states $|n,l>$ the level shifts $
\epsilon $ and widths $ \Gamma $ are quite accurately given by
\begin{equation} \label{g4}
\epsilon - i \Gamma/2 \simeq <n,l| V^{opt} | n,l>.
\end{equation}
For such average  values   the gradient potential $ V_G $ may be
expressed by a simpler formula  \cite{GRE87}
\begin{equation}
\label{g5}
 <n,l| \hat{V_G} | n,l>= \frac{2\pi}{\mu_{N N}} \frac{3}{4} a_P
\int d{\bf u } g_P({\bf u})\int d{\bf r}
 \rho({\bf r}-{\bf u}) \bar{D}^2 \Psi_{n,l}^{2}({\bf r}) +
\mid \Psi_{n,l} ({\bf r}-{\bf u} ) \mid^2 \Sigma_{\alpha}  D^2
\varphi_{\alpha}^{2}({\bf r}) +  V^{mix}
\end{equation}
where  $ \rho = \Sigma \mid {\varphi_{\alpha}}\mid^2 $ is the
nuclear density, $\Psi_{ n,l}$ is the atomic wave function and
\begin{equation}
\label{g6} \bar{D}^2 \Psi_{n,l}^{2}({\bf r})=
[\Psi_{r,n,l}^{'2}(r) + \frac{l(l+1)}{r^2} \Psi_{r,n,l}^{2}(r)]
\frac{1}{4\pi},
\end{equation}
\begin{equation}
\label{g7}
 D^2 \varphi_{\alpha}^{2}({\bf r})=[\varphi_{r,\alpha}^{'2}(r) +
\frac{L(L+1)}{r^2}\varphi_{r,\alpha}^{2}(r)] \frac{1}{4\pi}.
\end{equation}
In these equations $ \Psi_{r, n,l}$ denotes the radial part of the
atomic wave functions in states of main quantum numbers $n$ and
angular momentum $l$. The nucleon angular momentum is denoted by
$L$. Eqs.(\ref{g6}) and (\ref{g7}) are obtained with the "gradient
formula" which splits the derivative into tangential and radial
components. The two terms in eq.(\ref{O3}) which contain mixed
$\overrightarrow{\nabla}_{\bar{p}} \overrightarrow{\nabla}_{N}$
gradients generate $ V^{mix} $.  The latter may be expressed by a
formula obtained in Ref.~\cite{GRE87},
\begin{equation}
\label{g8}
 <\hat{V}^{mix}>  = \frac{2\pi}{\mu_{N N}}\frac{3}{4} a_P \int d{\bf x} \int d{\bf y}
[\overrightarrow{\nabla}\Psi_x^{*} \overrightarrow{\nabla}\Psi_x
g_{xy} \rho_y  + ( \Psi_x^{*} \frac {\Delta}{2} \Psi_x^{*}  +
\frac {\Delta}{2} \Psi_x^{*} \Psi_x)
 g_{xy} \rho_y   + (2i \mu_{N\bar{N}})^2
   \overrightarrow{j_x} g_{xy}\overrightarrow{j_y} ]
\end{equation}
A shorthand notation is used and arguments were put into indices
i.e. $ g_{xy} = g(\bf x -\bf y)$, $\Psi_x = \Psi(\bf x)$ etc. The
first and the second term in this equation tend to cancel
strongly. The last term, where $\overrightarrow{j} $ denotes the
nucleon and antiproton currents yields the main effect. For even A
(spin zero) nuclei the tangential currents average to zero and one is
left with the radial components which generate $ \Psi_{r,n,l}
\Psi_{r,n,l}^{'} \varphi_{r,\alpha}^{'} \varphi_{r,\alpha}$
contributions. However, in odd-A nuclei the tangential components in
the last term introduce correlations of the atomic and nuclear
currents. It leads to splitting of the atomic levels. Usually, this
effect is small, but it may be magnified if both antiproton and the
odd valence nucleon have high angular momenta.

Now some approximations are introduced to make the gradient potential
applicable to practical calculations.  For nucleons, the radial
gradients are needed only at large distances. In this region one has,
on average, $ \varphi_{r}^{'} \approx \varphi_{r}/2a $ where $a$ is
the surface thickness parameter. To account for the tangential
gradient, the angular momentum factor $L(L+1)$ in eq.(\ref{g6}) is
averaged over the three uppermost nucleon shells given by simple shell
model~\cite{BOH80}. The gradients of atomic wave functions may be
given explicitly for the circular orbits of interest. These functions
are
\begin{equation}
\label{g9}
 \Psi_{r,n,l}( r)= N(n,l) r^l \exp( -r/Bn) F_{nucl}(r),
\end{equation}
where N is a normalization, $B$ is the Bohr radius and $F_{nucl}$
describes the deformation of the Coulomb wave function due to short
range (nuclear + finite charge) interactions. The dominant effect in
$F$ comes from the damping due to absorption. For the radial
derivative one has $\Psi_{r,n,l}^{'} = ( l /r -1/nB + F'/F)
\Psi_{r,n,l}$ and the last term is calculated in a quasi-classical way
in terms of $ p_{WKB}(V) = \sqrt{2M_{\bar{N}}(E-V)}$ the value of
local momentum of the antiproton inside the nucleus. One has $ F'/F=
ip_{WKB}(V_{centr}+V_{cul}) -i p_{WKB}( V_{centr}+ V_{cul}+V^{opt})$
and numerically one finds that dropping the $F'/F$ term altogether
yields a minute change of the best fit parameters, making an overall
$\chi^2$ worse by 2.5. Also, it was found that a numerical calculation
of $\partial \Psi/\partial r$ makes no substantial improvement.

The approximation (\ref{g5}) is well fulfilled in the "upper" atomic
orbits.  Thus equation (\ref{g7}) is used here to $define$ the local
equivalent to the nonlocal gradient $V_G $ potential that gives the
same expectation value in a given atomic orbital of a given atom.
Expressions (\ref{g5},\ref{g6},\ref{g7}) lead to a typical folded
optical potential
\begin{equation}
\label{g10}
  \hat{V}_P(r) = \frac{2\pi}{\mu_{N N}}a_P\frac{3}{4}
\int d{\bf u} f_P({\bf u})D_r^2 \rho({\bf r}-{\bf u})
\end{equation}
where the effect of  antiproton and nucleon  momenta is  now
included into a function
\begin{equation}
\label{g11}
 D_r^2 =  [ \mid \frac{l}{r}  - \frac{1}{nB} + F'/F \mid^2 + \frac{l(l+1)}{r^2}] +
          [ \frac{1}{4 a^2}   + \frac{<L(L+1)>}{r^2}]
          + Re (\frac{l}{r}-\frac{1}{nB} + F'/F )\frac{1}{a}
\end{equation}
These approximations bring the optical potential back to the form
given by the basic equation \ref{O1} in the main text. Now $\hat{V}_P$
is given by eq.\ref{g10} summed over protons and neutrons. The
interpretation of three terms in eq.\ref{g11} is fairly
transparent. The first piece contains tangential and radial momenta of
the orbital antiproton.  It is the "localized" version of the gradient
term used in mesonic atoms. The second term contains radial and
tangential components of the nucleon momenta.  The radial ones are
expressed in terms of the nuclear density diffuseness parameter. The
last piece is the mixed term which contains only the radial term as
for spin zero systems the average product of tangential momenta
vanishes. All together the nucleon momentum part contributes about one
quarter of $V^{opt}$. On a phenomenological level it may be
approximately included into the $V_S$ potential term. The difference
arises when one attempts to relate $a_S $ and $a_P$ to the scattering
data. Another special effect of this nucleon term is due to its
dependence on the nucleon angular momenta. Some enhancement arises in
high $L$ shells e.g. in Pb nucleus ($12h$ protons and $14i $ neutrons
in the valence shell). Unfortunately, the $L(L+1)$ averaging procedure
contains some model dependence but the effect is moderate anyway,
since the $<L(L+1)>$ term constitutes less than $10\%$ of the total $
D_r^2$. In practical calculations we used a smooth, approximate
interpolation $ <L(L+1)> = 2+ (3Z/2-10 )/4$.  Rather slow but
systematic increase of $ D_r^2$ follows the rising atomic number $Z$.
It is due to the increase in the $L$ and $l$ values. The impact of the
nucleon $ <L(L+1)> $ term is also rather limited. Dropping it requires
a change of $a_P$ from the value $0.0-i 0.41 fm^3 $ (line 3 in Table
III) to the value $0.0-i 0.43 fm^3 $ and the $\chi^2$ increases by 4
units.

\acknowledgments

We are grateful to Peter Ring for his interest in this work and for
providing several nuclear density profiles. This work was supported by
the Deutsche Forschungsgemainschaft Bonn grant 436POL/17/8/04, and
Polish Ministry of Science project 1 P03B 042 29

\end{document}